\documentclass[aps,prl,reprint,superscriptaddress,floatfix]{revtex4-2}

\usepackage{color}
\usepackage{mathrsfs}
\usepackage{amsmath}
\usepackage{amssymb}
\usepackage{graphicx}
\usepackage[a4paper,top=2.00cm,bottom=2.00cm,left=2.00cm,right=2.00cm]{geometry}
\usepackage{color}
\usepackage{array}
\usepackage{multirow}
\usepackage{xspace}% correct spacing
\usepackage{bm}% bold math
\usepackage{xcolor}
\def\vp{{\bf p}}
\begin{document}

\title{Anisotropic superconductivity of niobium based on\\ its response to non-magnetic disorder}
\author{Makariy A. Tanatar}
\affiliation{Ames National Laboratory, Ames, IA 50011, USA}
\affiliation{Department of Physics \& Astronomy, Iowa State University, Ames, IA 50011, USA}

\author{Daniele Torsello}
\affiliation{Department of Applied Science and Technology, Politecnico di Torino, I-10129 Torino, Italy. }
\affiliation{Istituto Nazionale di Fisica Nucleare, Sezione di Torino, I-10125 Torino, Italy.}

\author{Kamal R. Joshi}
\affiliation{Ames National Laboratory, Ames, IA 50011, USA}
\affiliation{Department of Physics \& Astronomy, Iowa State University, Ames, IA 50011, USA}

\author{Sunil Ghimire}
\affiliation{Ames National Laboratory, Ames, IA 50011, USA}
\affiliation{Department of Physics \& Astronomy, Iowa State University, Ames, IA 50011, USA}

\author{Cameron J. Kopas}
\affiliation{Rigetti Computing, 775 Heinz Ave., Berkeley, CA 94710, USA}

\author{Jayss Marshall}
\affiliation{Rigetti Computing, 775 Heinz Ave., Berkeley, CA 94710, USA}

\author{Josh Y. Mutus}
\affiliation{Rigetti Computing, 775 Heinz Ave., Berkeley, CA 94710, USA}

\author{Gianluca Ghigo}
\affiliation{Department of Applied Science and Technology, Politecnico di Torino, I-10129 Torino, Italy. }
\affiliation{Istituto Nazionale di Fisica Nucleare, Sezione di Torino, I-10125 Torino, Italy.}

\author{Mehdi Zarea}
\affiliation{Center for Applied Physics and Superconducting Technologies,
             Department of Physics and Astronomy, Northwestern University,
	     Evanston, IL 60208, USA}

\author{James A. Sauls}
\affiliation{Hearne Institute of Theoretical Physics,
             Department of Physics and Astronomy, Louisiana State University,
             Baton Rouge, LA 70808, USA}
\affiliation{Center for Applied Physics and Superconducting Technologies,
             Department of Physics and Astronomy, Northwestern University,
             Evanston, IL 60208, USA}

\author{Ruslan Prozorov}
\email[Corresponding author: ]{prozorov@ameslab.gov}
\affiliation{Ames National Laboratory, Ames, IA 50011, USA}
\affiliation{Department of Physics \& Astronomy, Iowa State University, Ames, IA 50011, USA}

%\date{\today}
\date{29 August 2022}

\begin{abstract}
Niobium is one of the most studied superconductors, both theoretically and experimentally. It is tremendously important for applications, and it has the highest superconducting transition temperature, $T_{c}=9.33$ K, of all pure metals. In addition to power applications in alloys, pure niobium is used for sensitive magneto-sensing, radio-frequency cavities, and, more recently, as circuit metallization layers in superconducting qubits. A detailed understanding of its electronic and superconducting structure, especially its normal and superconducting state anisotropies, is crucial for mitigating the loss of quantum coherence in such devices. Recently, a microscopic theory of the anisotropic properties of niobium with the disorder was put forward. To verify theoretical predictions, we studied the effect of disorder produced by 3.5 MeV proton irradiation of thin Nb films grown by the same team and using the same protocols as those used in transmon qubits. By measuring the superconducting transition temperature and upper critical fields, we show a clear suppression of $T_{c}$ by potential (non-magnetic) scattering, which is directly related to the anisotropic order parameter. We obtain a very close quantitative agreement between the theory and the experiment.
\end{abstract}
\maketitle

\section{Introduction}

Niobium in its elemental form is an important material for modern technologies, from ultra-high quality factor superconducting microwave cavities \cite{Gurevich2012,ValenteFeliciano2022}, to superconducting circuits for sensitive magneto-sensing \cite{Clarke2004}, to applications in quantum information \cite{Reagor2016}. While the anisotropy of electronic and phononic band-structures of niobium was recognized a long time ago \cite{Ohta1978,Butler1980,Daams1980}, anisotropy of the superconducting order parameter received less attention \cite{Daams1980}. Recently, a self-consistent microscopic theory describing the anisotropic normal and superconducting states of niobium was put forward \cite{Zarea2022}.

Why is electronic anisotropy relevant and important for applications? There are many types of defects in solids \cite{Henderson1972,Tilley2008,Cai2016}, some are more, and some are less detrimental to the superconducting properties. Extended defects, such as dislocations, disclinations, stacking faults, and grain boundaries, mostly affect the macroscopic supercurrent flow without affecting the order parameter in the bulk, such as local superconducting transition temperature, $T_c$, and the value of the superconducting order parameter. Point-like defects, on the other hand, exist in the whole volume, and they interact with the Cooper pairs everywhere.

As far as electron scattering on defects is concerned, there are two different types of point-like defects in crystals. Scattering on potential, a.k.a. ``non-magnetic", defects involves only the Coulomb interaction with conduction electron regardless of the spin value and state. In superconductors with isotropic $s-$wave order parameter, such ``non-magnetic" scattering does not change $T_c$. This statement is known as Anderson's theorem \cite{Anderson1959a}. This is true irrespective of the anisotropy of the Fermi surface. The second type of defect scatters by flipping its spin and simultaneously flipping the spin of the scattered conduction electron. If a scattered electron was part of a spin-singlet Cooper pair, the pair will be broken. Since the order parameter magnitude (and hence $T_c$) depend on the total number of Cooper pairs,  magnetic impurity scattering will reduce these quantities. This was shown by Abrikosov and Gor'kov in 1960 \cite{AbrikosovGorkov1960ZETF}. The situation becomes more subtle if the order parameter is anisotropic, the material is a multiband metal with different gaps for different bands, or both. Generally, in such cases both types of defects are pair-breaking, although to a different degree \cite{Efremov2011,Golubov1997,Hirschfeld1993PRB,Kogan2016,Openov2004,Openov1997,Torsello2019JOSC,Zarea2022}.

If niobium has significant anisotropy of its superconducting state, then special care should be taken to avoid all kinds of defects unless introduced deliberately for some reason, such as the enhancement of the Ginzburg-Landau parameter, $\kappa$. Here we report on the effects of non-magnetic defects induced by 3.5 MeV proton irradiation on the superconducting properties of a 160 nm niobium film used in the fabrication of transmon qubits.

It is quite difficult to study the anisotropy experimentally in a highly symmetric body-centered cubic metal. In the superconducting state, this is further complicated by non-ideal shapes of real samples, leaving only a limited selection of properties to be probed. Traditionally, it was the upper critical field, $H_{c2}$, measured along the {[}100{]}, {[}110{]} and {[}111{]} directions \cite{Finnemore1966,Butler1980,Arai2004}.  However, the presence of disorder significantly affects the measurements and smears the anisotropy \cite{HW1966,Kogan2013}. More importantly, though, is that the anisotropy of $H_{c2}$ is mostly determined by the anisotropic Fermi velocity, $H_{c2}\sim v^{-2}$ \cite{HW1966} and not by the superconducting order parameter (OP), which is our main interest here. Directional tunneling is the direct probe of the density of states, but it is very sensitive to the surface quality and additional layers usually formed on fresh niobium surface, such as niobium oxides \cite{Lechner2020}. An additional complication in the case of niobium is that the predicted direct and reciprocal space distribution of significantly anisotropic variation of the OP is confined to fairly narrow angular intervals along the principle directions \cite{Zarea2022}, which makes directional measurements more difficult.

Another approach to studying the anisotropy of the order parameter utilizes the sensitivity of the superconducting transition temperature, $T_{c}$, and of the upper critical field, $H_{c2}$, to disorder scattering. As explained in more detail above, there are four possible scenarios. (1) Isotropic OP and potential (non-magnetic) disorder. In this case, $T_{c}$ does not change, as described by the so-called Anderson theorem \cite{Anderson1959a}.  We note it is true for \emph{any} anisotropic Fermi surface \cite{Openov2004}, but generally does not hold for a multi-band superconductor. $H_{c2}$ \emph{increases} almost linearly proportional to the scattering rate, $\Gamma=\hbar\left(2\pi k_{B}T_{c0}\tau\right)^{-1}$, where $T_{c0}$ is the initial transition temperature and $\tau$ is the characteristic scattering time \cite{HW1966,Kogan2013,Gurevich2003,Xie2017}. (2) Isotropic OP with magnetic (spin-flip) disorder scattering follows the Abrikosov-Gor'kov theory \cite{AbrikosovGorkov1960ZETF}. Here, $T_{c}$ can be suppressed all the way to zero at the finite scattering rate, $\Gamma=0.14$. However, opposite to the previous case, $H_{c2}$ \emph{decreases} with this pair-breaking scattering \cite{Kogan2013}.  (3,4) Anisotropic OP (hence $T_{c}$) is suppressed by both magnetic and non-magnetic impurities~\cite{Hirschfeld1993PRB,Openov2004,Openov1997,Golubov1997} and this can be readily extended to the multiband superconductors \cite{Kogan2009PRB-2,Efremov2011,Torsello2019JOSC}. The degree of suppression depends sensitively on the anisotropy of the order parameter and may or may not drive the $T_{c}$ all the way to zero. The upper critical field behaves similarly to cases 1 and 2, increasing with the potential scattering and decreasing with spin-flip scattering \cite{Kogan2013}.  Therefore, the radiation-induced disorder seems to be an ideal way to study OP anisotropy. However, one has to be careful not to alter the electronic structure and not to dope the material because, among other parameters, $T_{c}$ depends sensitively on the density of states at the Fermi level. The same should be said regarding the phonon spectra. Obviously, chemical doping with other elements, as it is often done, is not the cleanest way to produce controlled disorder.

The nature of the disorder induced by particle irradiation depends on the particles used as well as irradiation conditions, particularly temperature. Different irradiated compounds respond differently, therefore each case should be considered individually. In the case of niobium, annealing of the induced defects judged by the decrease of the residual resistivity was studied after 4.5 K low-temperature 3 MeV electron irradiation where practically complete annealing was found by the time the sample reached room temperature \cite{Faber1977}. A similar trend was observed in a 4.6 K, neutron irradiation experiment \cite{Berndt1970}. On the other hand, neutron irradiation conducted at room temperature resulted in a meta-stable population of defects that could only be annealed at higher temperatures \cite{Peacock1963}. Therefore, it seems that room-temperature irradiation of niobium is more effective than the low-temperature one after which the sample is brought to room temperature and then cooled and measured. Indeed, if the sample is not warmed up after low-temperature irradiation, a substantial amount of defects is created. When $T_c$, $\rho_0$ and $H_{c2}$ were measured after 4.6 K neutron irradiation without warming to higher temperatures, a monotonic decrease of $T_c$ vs. $\rho_0$ was found, pointing to the anisotropic superconducting gap in niobium \cite{Berndt1970}. The upper critical field increased confirming the increase of potential scattering with irradiation. \cite{Berndt1970}. Presented here proton irradiation was performed at room temperature and resulted in a clear increase of the residual resistivity and $H_{c2}$, and a consistent change of $T_c$ that quantitatively fits well the recent microscopic theory \cite{Zarea2022}.

%----------- Figure 1 -----------
\begin{figure}[tb]
\centering \includegraphics[width=6cm]{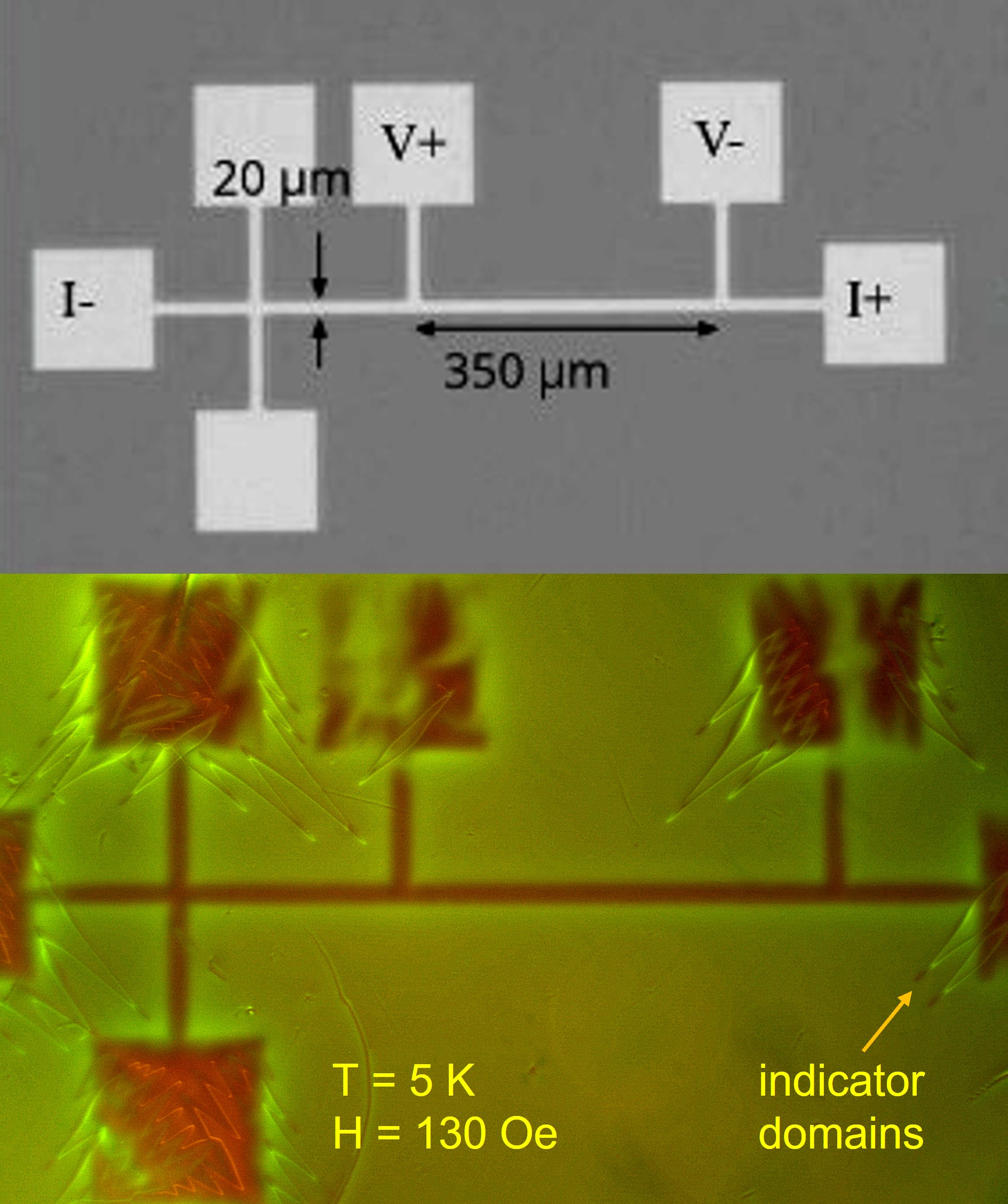}
\caption{(color online) Top frame: optical image of a bridge structure made of a 160 nm niobium film sputtered on a [001] silicone substrate. The in-plane dimensions of the bridge are shown. Lower frame: magneto-optical image at 5 K showing excellent shielding of 130 Oe of the applied magnetic field, indicating very good connectivity and material homogeneity in the structure.}
\label{fig1}
\end{figure}
%---------------------------------

In the past few decades, artificial point-like disorder induced by electron and proton irradiation emerged as a powerful tool to probe the superconducting state, in particular, the anisotropy of the superconducting order parameter via the measurements of $T_{c}$ and London penetration depth \cite{Cho2018SST_review,FeRh122_PRL2018,Torsello2020PrAppl}. In the case of niobium, a known conventional spin-singlet superconductor, the latter is not particularly needed since exponential attenuation is expected in clean and dirty limits, but the $T_{c}$ and $H_{c2}$ can be studied as experimental parameters sensitive to both types of disorder, magnetic and non-magnetic.
We use proton irradiation to modify a niobium film deposited on a silicon substrate, similar to films used in many superconducting qubits \cite{Gambetta2017, Nersisyan2019, Premkumar2021}. The results support the recent theory \cite{Zarea2022} of anisotropic superconductivity in metallic niobium.

%---------------- Figure 2 ------------------
\begin{figure}[tb]
\centering \includegraphics[width=8cm]{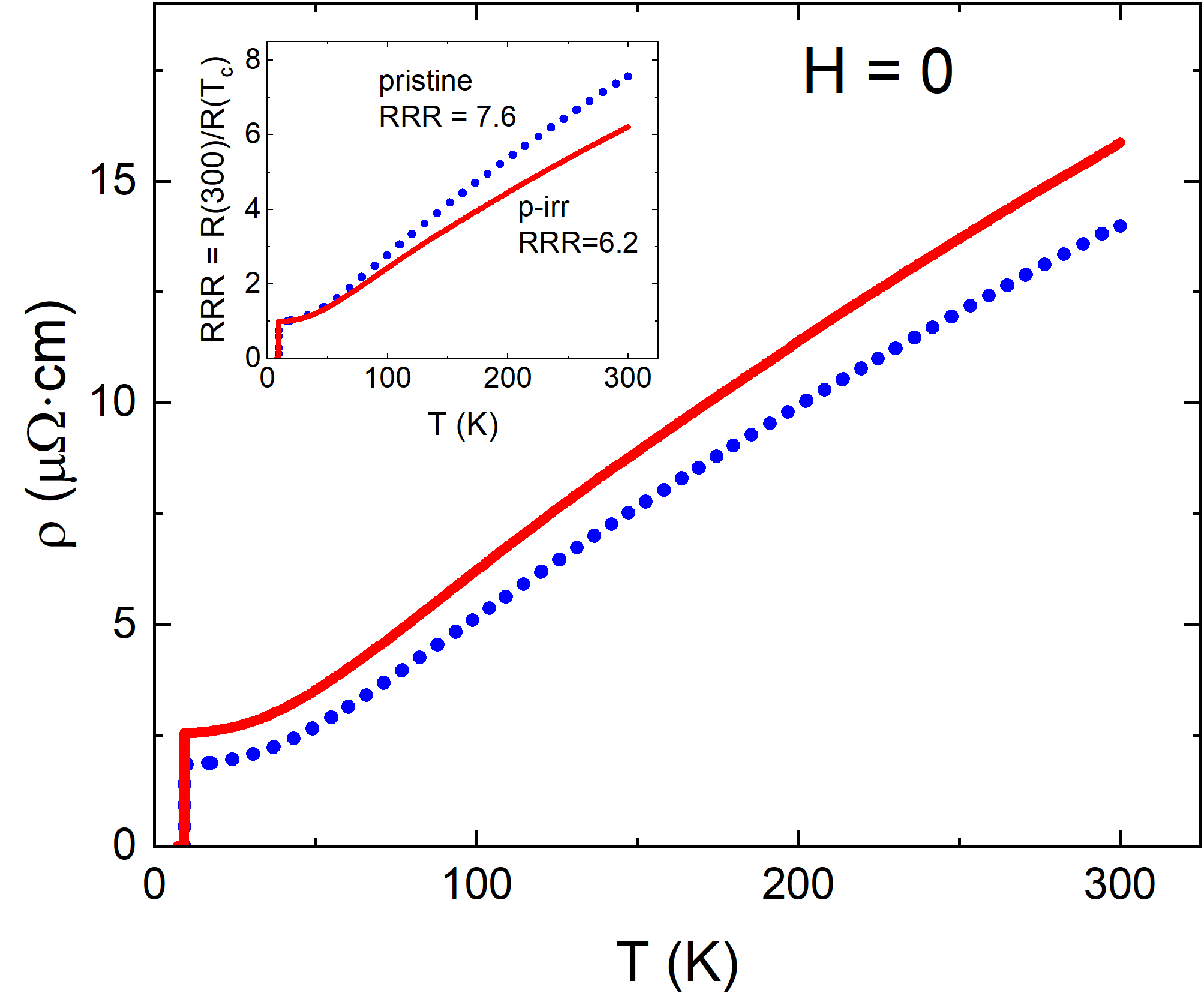}
\caption{(color online) Resistivity of a 160 nm thick niobium film on a silicon substrate measured between room temperature and below the superconducting transition.  The blue dotted line is for the pristine state and the solid red line is for the same film after proton irradiation that created $D=2.4\times10^{-4}$ defects per atom (dpa) corresponding to roughly one defect per 2000 BCC unit cells. Inset shows the resistivity normalized by its values at $T_{c}.$ This gives the residual resistivity ratio, $RRR$ decreasing from 7.6 in the pristine sample to 6.7 after the irradiation.}
\label{fig2}
\end{figure}
%--------------------------------------------------

\section{Experimental}

Niobium films were deposited using high power impulse magnetron sputter deposition to $\sim$160 nm thick, onto high resistivity ($\geq 10000$ $\Omega \cdot$cm) undoped  {[}001{]} Si substrates.  Test structures were patterned using standard photo-lithography techniques into a bridge structure suitable for accurate 4-point resistivity measurements. A picture of the resistivity test sample with dimensions is shown in the upper panel of Fig.\ref{fig1}.

The quality of the bridge structure was examined using magneto-optical imaging performed in a closed-cycle ﬂow-type optical $^4$He cryostat using Faraday rotation of polarized light in bismuth-doped iron-garnet ﬁlms with in-plane magnetization \cite{Prozorov2008}. In the bottom panel of Fig.\ref{fig1} the intensity is proportional to the local magnetic induction. The dark area of the bridge shows a perfect shielding of 130 Oe magnetic field applied at 5 K after cooling the structure in zero field. The zigzag structure is from the in-plane magnetic domains in the Faraday indicator and does not affect the result.

Proton irradiation was performed at the van der Graaf - type CN proton accelerator of the Legnaro National Laboratory (LNL) of the Italian National Institute for Nuclear Physics (INFN). The irradiation was carried out at room temperature in a high vacuum with a 3.5 MeV defocused proton beam perpendicular to the surface of the sample.  At the fluence of $\Phi=6\times10^{16}$ protons/cm$^{2}$, SRIM (The Stopping and Range of Ions in Matter) \cite{Ziegler2010} calculations show that the sample has acquired an irradiation dose that produced $D=2.4\times10^{-3}$ dpa (defects per atom), which is roughly one defect per $200$ BCC unit cells of Nb. Of course, a portion of the generated defects, which are mostly Frenkel pairs of vacancy-interstitial, will recombine, and their population relaxes \cite{Torsello2021SciRep}. However, we do not use the calculated defect number and only the measured resistivity and temperature. SRIM calculations of proton penetration into Nb film on Si substrate show the peak of the energy deposition and the implantation peak deep inside the substrate, at about 120 $\mu\mathrm{m}$ from the Nb film. This is too far for protons to migrate back to the film. However, if they do migrate and even form niobium hydride, it will have nanoscale nature as it was recently shown on similar films \cite{Lee2021}. They will also play the role of the scattering centers. For the analysis, we only need to know a measurable change of resistivity, proportional to the number of defects, and the change of the transition temperature that is directly connected to the gap anisotropy.

Four probe electrical resistivity measurements were performed in {\it Quantum Design} PPMS. Measurements were performed on the same bridge structure before and after proton irradiation with $D=$2.4$\times 10^{-4}$ dpa. Contacts to the contact pads were created by gluing 25 $\mu$m silver wires using DuPont 4929N conducting silver paste. This technique provides contacts with contact resistance in  10 to 100 $\Omega$ range. Additionally, resistivity measurements were also performed on non-patterned Nb film on Silicon substrate, grown in the same conditions as the bridge structure. Contacts to both films were removed for irradiation, so the geometric factor of the samples was different for measurements before and after irradiation. Both the bridge and the unpatterned film were subjected to the same irradiation dose. They showed identical $T_c$ in the pristine state and identical suppression after irradiation.

Transport measurements of the upper critical field, $H_{c2}$, were performed with a magnetic field oriented perpendicular to the film plane to avoid the third critical field appearing in the case when a magnetic field has the component parallel to the film surface \cite{Abrikosov2017}.

\section{Results}

%--------------------------- Figure 3 --------------------------
\begin{figure}[tb]
\centering \includegraphics[width=8cm]{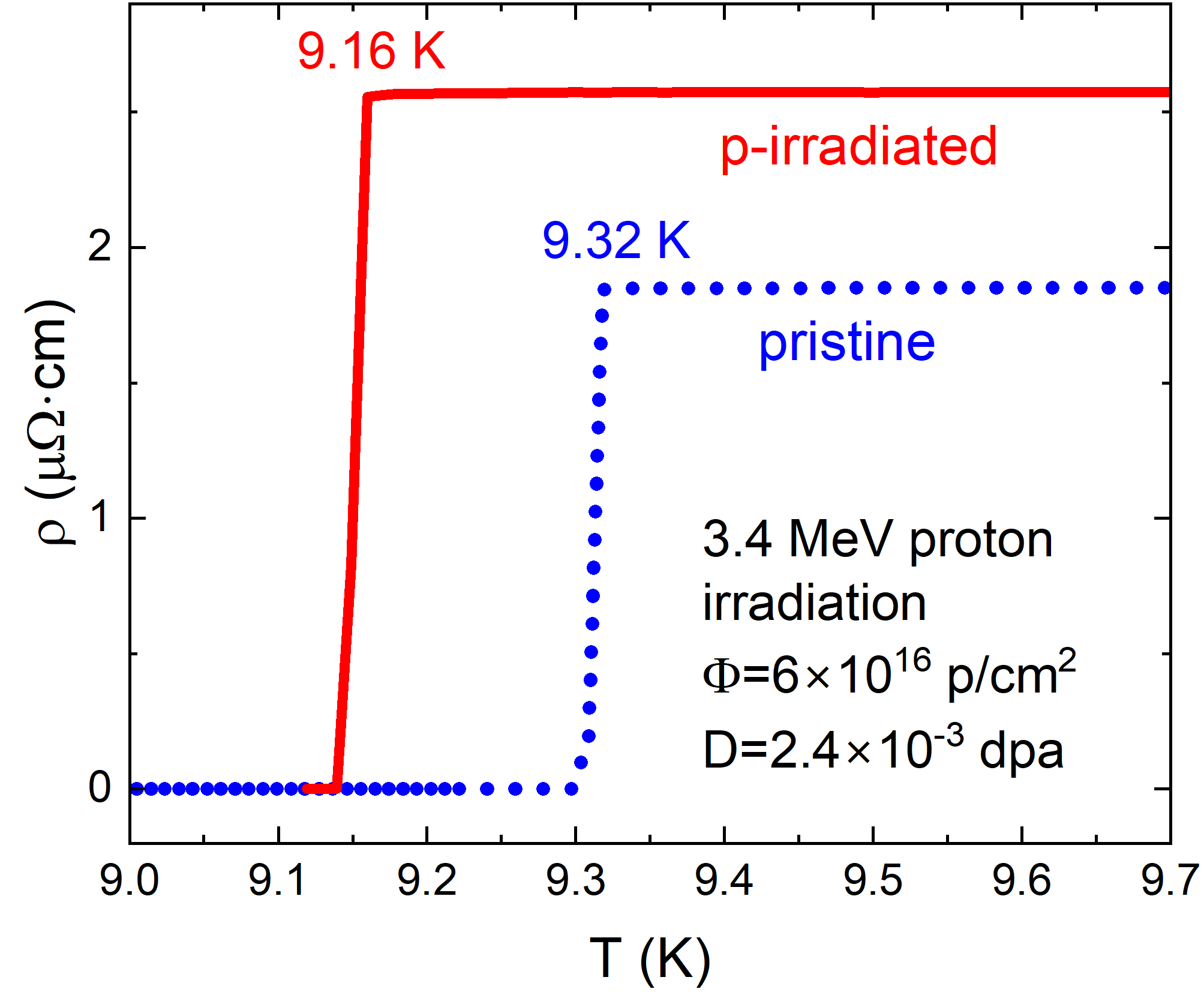}
\caption{(color online) Resistivity of 160 nm niobium film on a silicon substrate in the vicinity of the superconducting transition, $T_{c}$. The blue dotted line shows the pristine state, and the solid red line is the same film after proton irradiation. The superconducting transition temperature shifts by, $\Delta T_{c}=9.33-9.16=0.17$ K, while the resistivity at the transition changed by $\Delta\rho\left(T_{c}\right)=2.59-1.87=0.72\:\mathrm{\mu\Omega\cdotp cm}$.}
\label{fig3}
\end{figure}
%----------------------------------------------------------

Figure \ref{fig2} shows temperature-dependent resistivity of 160 nm niobium film on a silicon substrate before (dotted blue line), and after (solid red line) proton irradiation, measured in zero applied magnetic field from room temperature to below the superconducting transition. The entire $R(T)$ curve is shifted up after the irradiation, proportional to the additional scattering introduced. A slight increase in the slope of $\rho (T)$ curve after irradiation is most likely due to geometric factor change. The inset shows the normalized resistivity, $RRR=R(300K)/R(T_{c})$. This gives the residual resistivity ratio, $RRR$ decreasing from 7.6 in the pristine sample to 6.7 after the irradiation.

Figure \ref{fig3} zooms into the superconducting transition. The result is clear - the transitions remains equally sharp after the irradiation (indicating that no inhomogeneity has been introduced), but $T_c$ shifts down from $T_{c0}=9.33$ K to $T_{c}=9.16$ K, $\Delta T_{c}=0.17$ K, while the resistivity at the transition changed by $\Delta\rho\left(T_{c}\right)=2.59-1.87=0.72\:\mathrm{\mu\Omega\cdotp cm}$.  These changes may seem insignificant, but they are clearly resolvable.  We note that we observed similar trends in several other samples of niobium irradiated at similar and different doses.

%--------------------------- Figure 4 -----------------------------
\begin{figure}[tb]
\centering \includegraphics[width=8cm]{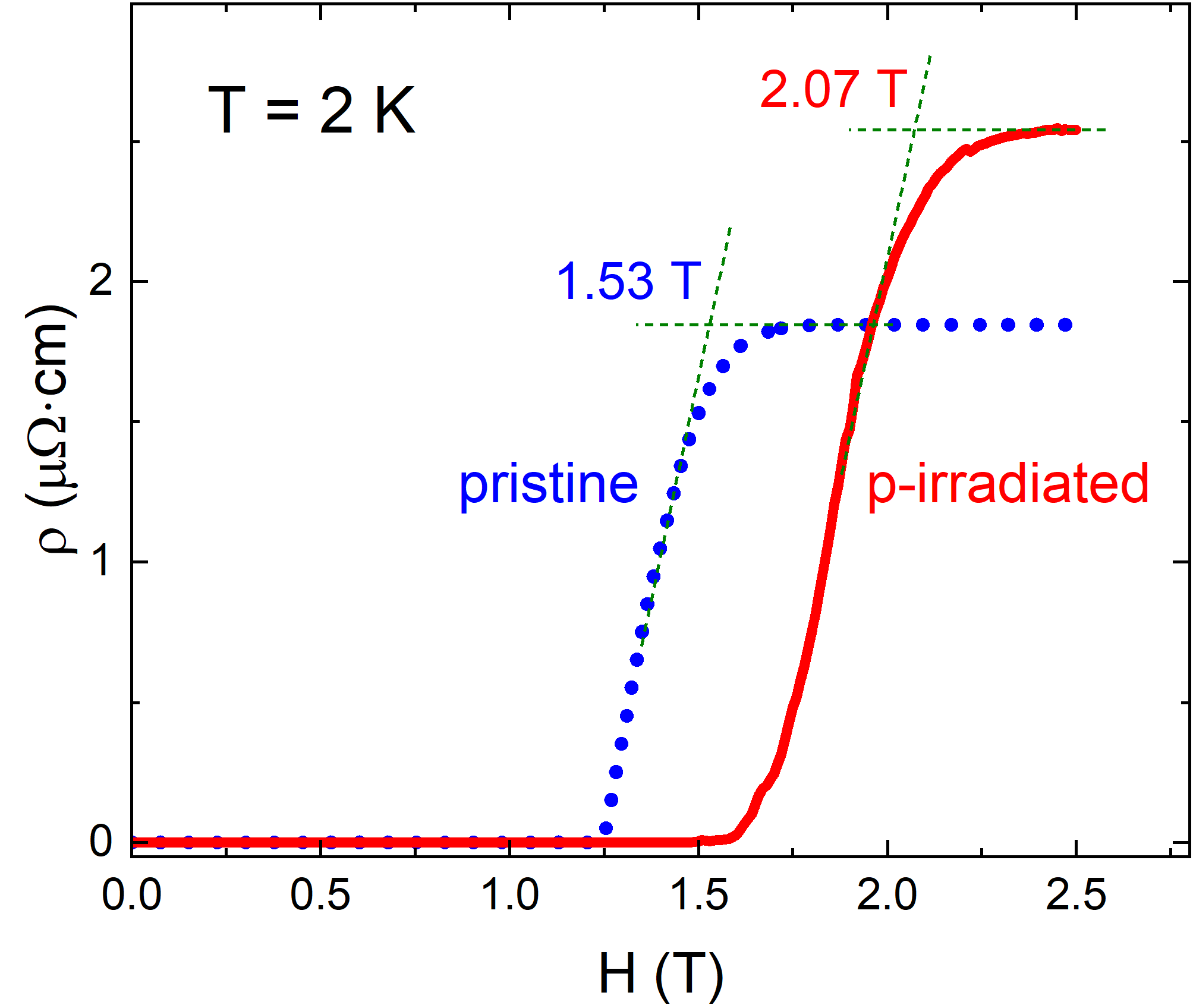}
\caption{(color online) Resistivity at $T=2$ K as a function of a magnetic field before (dotted blue line) and after (solid red line) the proton irradiation.  The upper critical field increases from $1.53$ T before irradiation to $2.07$ T after.}
\label{fig4}
\end{figure}
%--------------------------------

So far, we have established that defects produced by proton irradiation suppress the transition temperature $T_{c}$. As described in the introduction, if these defects were magnetic, they would suppress $T_{c}$ regardless of the order parameter anisotropy. We, therefore, need to establish the nature of the induced defects. Figure~\ref{fig4} shows magnetic field dependence of resistivity at $T=2$ K before (dotted blue line) and after (solid red line) proton irradiation. The upper critical field increases from $1.53$ T before irradiation to $2.07$ T after, the enhancement by a factor of $1.35$. This clearly means that we are dealing with non-magnetic impurities and, therefore, the suppression of $T_{c}$ is solely due to the anisotropy of the superconducting order parameter. It also excludes possible damage and deterioration of our sample because in such case, the $H_{c2}$ would either remain unchanged or decrease.

\section{Comparison with Theory}

In order to compare the experimental results obtained for the suppression of $T_c$ by radiation-induced disorder with theoretical predictions based on non-magnetic scattering and gap anisotropy \cite{Zarea2022}, we first determine the scattering rate of electrons and holes by the random potential. The electron-impurity scattering rate is determined from the linear dependence of the upper critical field at low temperature with the dimensionless parameter, $\Gamma\equiv\hbar/2\pi\tau k_{\mbox{\tiny B}} T_{c_0}$. N.B. $\Gamma$ is the scattering rate $1/\tau$ times the pair formation time, $t_{\mbox{\tiny coh}}=\hbar/2\pi k_{\mbox{\tiny B}}T_{c_0}$.

The upper critical field is related to the pair correlation length, $\xi$, by
\begin{equation}
H_{c_2} = \frac{\Phi_0}{2\pi^2\xi^2}
\,,
\end{equation}
where $\Phi_0=hc/2e$ is the flux quantum.
For an isotropic superconductor in the clean limit ballistic propagation at the Fermi velocity generates a pair correlation length
\begin{equation}
\xi_{\mbox{\tiny ballistic}}=v_f\,t_{\mbox{\tiny coh}}=\hbar v_f/2\pi k_{\mbox{\tiny B}}T_{c}\equiv\xi_0
\,,
\end{equation}
and thus an upper critical field of $H_{c_2}^0=\Phi_0/2\pi\xi_0^2$.

For isotropic (``s-wave'') pairing the transition temperature, and thus the pair formation time, $t_{\mbox{\tiny coh}}$, is insensitive to non-magnetic disorder.
However, disorder disrupts ballistic propagation, and thus the spatial correlation length will decrease for finite mean free path, $\ell=v_f\tau$. In the limit $\ell \ll \xi_0$ the spatial scale of pair formation is determined by \emph{diffusive} transport of electrons. Thus, spatial pair correlations are governed by diffusion on the timescale for pair formation. The diffusion propagator in three spatial dimensions is given by
$G({\bf r},t)=(4\pi {\mathcal D}t)^{-{\small 3/2}}\,e^{-r^2/4{\mathcal D}t}$,
where ${\mathcal D}=\frac{1}{3} v_f\,\ell$ is the diffusion constant for electrons moving with the Fermi velocity and scattering with mean free path $\ell=v_f\,\tau$. Thus, the pair correlation length in the diffusive limit is given by the leading edge of the diffusion front at time $t_{\mbox{coh}}$,
\begin{equation}
\xi_{\mbox{\tiny diffusive}}^2=4{\mathcal D}\, t_{\mbox{coh}}= \frac{4}{3}\,\xi_0^2\,\frac{1}{\Gamma}
\,.
\end{equation}
In the weak disorder limit, $\ell < \xi_0$, impurity scattering will suppress the pair correlation length perturbatively, giving rise to a monotonic cross-over from the ballistic result and the diffusive result. We capture this smooth evolution as a function of $\Gamma$ as
$1/\xi^2 = 1/\xi^2_{\mbox{\tiny ballistic}} + 1/\xi^2_{\mbox{\tiny diffusive}}=\xi_0^{-2}\left(1+\frac{3}{4}\Gamma\right)$,
and thus scaling of the upper critical field as

\begin{equation}
H_{c2} = H_{c2}^0\,\left(1 + \frac{3}{4} \Gamma\right)
\,,
\label{Hc2Gamma}
\end{equation}
a result that is born out by microscopic calculations for the upper critical field with non-magnetic disorder for isotropic superconductors~\cite{Kogan2013,Xie2017},

This simple, linear in $\Gamma$ behavior of $H_{c2}$, provides a natural way to estimate $\Gamma$ from the measured upper critical field. This is quite fortunate and particularly important in the case of thin films where, due to granularity, electrical resistivity cannot be used for a direct estimate of the scattering time. In our experiments, both $T_c$ and $H_{c2}$ were measured before and after the irradiation. Indeed, what we call a "pristine" sample does not imply $\Gamma=0$ - there are natural defects. In fact, there is quite a substantial scattering already.

In the experiment, $T_c$ decreased from 9.32 K to 9.16 K after proton irradiation, see Fig.\ref{fig3}. The upper critical field increased from 1.53 T to 2.07 T, Fig.\ref{fig4}. Assuming $H_{c2}(\Gamma=0) \equiv H_{c2}^0=0.5$~T obtained for the purest single crystals with RRR=15000 \cite{Finnemore1966}, we obtain the ratios $H_{c2}/H_{c2}^0=3.06$ and $4.14$ before and after the irradiation, respectively. Using Eq.\ref{Hc2Gamma} we estimate the increase of $\Gamma=4/3(H_{c2}/H_{c2}^0-1)$ from 2.75 before to 4.19 after the irradiation.

In the presence of gap anisotropy non-magnetic impurity scattering is pair-breaking, leading to violation of Anderson's theorem and a suppression of the superconducting transition. The magnitude of the suppression
depends on the quasiparticle-impurity scattering rate, $1/\tau$, and the RMS deviation of the gap from its mean via the normalized parameter,
\begin{equation}
{\mathcal A} \equiv \lim_{T\rightarrow T_c}
\frac{
\langle|\Delta(\vp)|^2\rangle
-
|\langle\Delta(\vp)\rangle|^2
}
{
\langle|\Delta(\vp)|^2\rangle
}
\,.
\end{equation}
Note that in the limit $T\rightarrow T_c$ the order parameter, $\Delta(\vp)=\Delta\,{\mathcal Y}(\vp)$, is proportional to the anisotropic Cooper pair amplitude, ${\mathcal Y}(\vp)$, in momentum space confined to the Fermi surface, and is the eigenfuction of the linearized gap equation belonging to the irreducible representation of the point group with eigenvalue determining the highest $T_c$.
The result for $T_c$ as a function of $\mathcal A$ and $1/\tau$ is the solution of the transcendental equation~\cite{Zarea2022},
\begin{equation}
\ln \frac{T_{c_0}}{T_c}={\mathcal A}\times
2\pi T_c\,\sum_{\varepsilon_n>0}^{\infty}
\left(
\frac{\hbar/\tau}
{\varepsilon_n(\varepsilon_n+\frac{1}{2}\hbar/\tau)}
\right)
\,,
\label{Tc_vs_A-tau}
\end{equation}
where $\varepsilon_n=\pi T_c(2n+1)$ are the Fermion Matsubara energies, and is a generalization of the result obtained Larkin for pair-breaking by non-magnetic scattering in superconductors with p-wave pairing~\cite{lar65} to any single-band anisotropic superconductor. Larkin's result for p-wave pairing - actually any unconventional superconductor with $\langle\Delta(\vp)\rangle\equiv 0$ - corresponds to $\mathcal A=1$. See also Refs.~\cite{thu96,thu98,Openov1997,Openov2004}. Anderson's theorem is recovered for an isotropic Cooper pair amplitude, $\mathcal Y = 1$, i.e. $\mathcal A = 0$.
The solution of Eq.~\eqref{Tc_vs_A-tau} yields $T_c=T_{c_0}\times S(\mathcal A,\Gamma)$, where $\Gamma=\hbar/\left(2\pi k_{B}T_{c_0}\tau\right)$ is the product of the scattering rate, $1/\tau$, and the Cooper pair formation time, $\hbar/2\pi k_{B} T_{c_0}$.
The function $S(\mathcal A,\Gamma)$ versus $\Gamma$ over the full parameter range of anisotropy $0\le\mathcal A \le 1$ is reported in Fig. (10) in Ref.~\cite{Zarea2022}. Note that $T_c$ does not converge to a non-zero value for $\Gamma\rightarrow\infty$, except for the exceptional case of $\mathcal A = 1$.
However, this ignores the physical cutoff set by the phonon bandwidth $\sim\Omega_D$ in the otherwise convergent sum in Eq.~\eqref{Tc_vs_A-tau}. In the strong disorder limit, $1/\tau > \Omega_D$ scattering by the random potential is sufficiently fast to average the Cooper pair amplitude over the Fermi surface, leading to a uniform gap amplitude on the Fermi surface and $T_c$ independent of disorder for $\Gamma \gtrsim 1/\Omega_D\,\tau$~\cite{rai86}.
For the Nb samples considered here we are in the moderate disorder limit governed by Eq.~\eqref{Tc_vs_A-tau}. For weak disorder, $\Gamma\ll 1$, the perturbative solution is $T_c \simeq T_{c_0}\,\left(1-\mathcal A\,\frac{\pi^2}{4}\,\Gamma\right)$, a result first obtained by Tsuneto, reported in Ref.~\cite{AGD}. However, this limit is rarely achieved. For intermediate disorder $\Gamma \lesssim 1$ and modest anisotropy, $\sqrt{\mathcal A} \lesssim 0.2$, we need the non-universal function $S(\mathcal A,\tau)$ to make accurate predictions and analysis. From DFT and Eliashberg theory Zarea et al. obtain $\mathcal{A}=0.037$~\cite{Zarea2022}.
In order to check the consistency of the experimental results with the predictions of the microscopic theory it is worth noting that the transiton temperature for pure, bulk single crystalline Niobium is poorly established. The value calculated in Ref.~\cite{Zarea2022} of $T_{c_0}=9.33\,\mbox{K}$ is obtained with rather large renormalized Coulomb repulsion, $\mu^*=0.22$, that suppresses the binding of Cooper pairs from the attractive electron-phonon interaction. Thus, the transition temperature of pure single-crystalline Nb may be higher than $9.33\,\mbox{K}$. This seems to be the case since the measured transition temperature of our Nb film {\it before} proton irradiation is $T_c=9.32\,\mbox{K}$, which is impossible to reconcile
with $T_{c_0}=9.33\,\mbox{K}$, $\Gamma=2.75$ and substantial anisotropy of the order parameter.
This is also consistent with the literature on Nb which shows a significant spread in reports for the value of $T_c$, including appreciably higher values, e.g. $T_c=9.7\,\mbox{K}$~\cite{Gubin2005}.
Thus, we assume $T_{c_0}$ is higher than that reported in Ref.~\cite{Zarea2022}, then analyze the suppression of $T_c$ for the film before and after proton irradiation based on Eq.~\ref{Tc_vs_A-tau} with the values of $\Gamma$ obtained from the effect of disorder on $H_{c_2}$.
Now obtain $T_{c_0}$ based on the anisotropy parameter $\mathcal{A}=0.037$ by requiring
$T_c(\mathcal A,\Gamma=2.75)=9.32\,\mbox{K}$ for the Nb film prior to irradiation. Equation~\eqref{Tc_vs_A-tau} then gives $S(\mathcal A,\Gamma)=0.91545$, and thus $T_{c_0}=9.32/0.91545=10.18\,\mbox{K}$.
Note that this is the theoretical limit for $\Gamma=0$, i.e. pure, single-crystalline Nb, and that the rate of suppression of $T_c$ from $T_{c_0}$ is largest in the limit $\Gamma\rightarrow 0$.
With this value for $T_{c_0}$ and the value $\Gamma=4.19$ for the Nb film after the irradiation Eq.~\eqref{Tc_vs_A-tau} yields $S(\mathcal A,\Gamma=4.19)=0.901096$, and thus predicts $T_c(\mathcal A,\Gamma=4.19)=10.18 \times 0.901096=9.174\,\mbox{K}$, which is in excellent agreement the measured value of $T_c=9.16\,\mbox{K}$ for the irradiated Nb film as shown in Fig.~\ref{fig3}.
In our view this agreement with the theory of pair-breaking by non-magnetic disorder for anisotropic pairing in Nb is impressive. However, we are led to the conclusion that the maximum transition temperature for pure, single crystalline is $T_{c_0}=10.18\,\mbox{K}$. As noted earlier, this may be compatible with Eliashberg theory if we can justify a lower value of the renormalized Coulomb interaction, an issue that will be addressed separately from this report. In any event the larger value for $T_{c_0}$ will not effect the anisotropy of the order parameter as that is determined by the momentum-dependent electron-phonon coupling.

In conclusion, 3.5 MeV proton irradiation was used to introduce non-magnetic disorder in a 160 nm niobium film. By measuring the transition temperature and the upper critical field before and after the irradiation, we conclude that the observed changes follow the recent microscopic theory predicting specific anisotropic order parameters closely. We introduced a novel way to estimate dimensionless scattering rate based on the upper critical field, rather than usually used resistivity measurements, which may suffer from additional inter-grain contributions, especially in sputtered films. We obtain a remarkably quantitative agreement between the experiment and the theory with only one parameter to vary - the theoretical transition temperature in clean material, $T_{c0}=10.18\,\mbox{K}$.

\section{Acknowledgments}
We thank Vladimir Kogan for fruitful discussions and the Rigetti Computing fabrication team for developing and manufacturing niobium films used in this work. This work was supported by the U.S. Department of Energy, Office of Science, National Quantum Information Science Research Centers, Superconducting Quantum Materials and Systems Center (SQMS) under contract number DE-AC02-07CH11359. The research was performed at the Ames Laboratory, which is operated for the U.S. DOE by Iowa State University under contract \# DE-AC02-07CH11358. The proton irradiation was performed at the CN facility of INFN-LNL. D.T. and G.G. were supported by the Italian Ministry of Education, University and Research (Project PRIN HIBiSCUS, Grant No. 201785KWLE), and by INFN CSN5, through the experiment SAMARA.

%-----------------------------
%\bibliographystyle{apsrev4-2}
%\bibliography{Nb-SQMS}

\end{document}